# Specific heat of $Ba_{0.59}K_{0.41}Fe_2As_2$, and a new method for identifying the electron contribution: two electron bands with different energy gaps in the superconducting state

C. R. Rotundu,[1,2,*,a,b] T. R. Forrest,[3,4,a,c] N. E. Phillips,[1,5] and R. J. Birgeneau[1,3,6]

[1] *Materials Sciences Division, Lawrence Berkeley National Laboratory, Berkeley, California 94720, USA*

[2] *Department of Physics and Astronomy, UCLA, Los Angeles, CA 90095, USA*

[3] *Department of Physics, University of California, Berkeley, California 94720, USA*

[4] *European Synchrotron Radiation Facility, BP 220, F-38043 Grenoble Cedex, France*

[5] *Department of Chemistry, University of California, Berkeley, California 94720, USA*

[6] *Department of Materials Science and Engineering, University of California, Berkeley, California 94720, USA*

We report measurements of the specific heat of $Ba_{0.59}K_{0.41}Fe_2As_2$, an Fe-pnictide superconductor with $T_c$ = 36.9 K, for which there are suggestions of an unusual electron pairing mechanism. We use a new method of analysis of the data to derive the parameters characteristic of the electron contribution. It is based on comparisons of $\alpha$-model expressions for the *electron contribution* with the *total* specific heat, which give the electron contribution *directly*. It obviates the need in the conventional analyses for an independent, necessarily approximate, determination of the lattice contribution, which is subtracted from the total specific heat to obtain the electron contribution. It eliminates the uncertainties and errors in the electron contribution that follow from the approximations in the determination of the lattice contribution. Our values of the parameters characteristic of the electron contribution differ significantly from those obtained in conventional analyses of specific-heat data for five similar hole-doped $BaFe_2As_2$ superconductors, which also differ significantly among themselves. For $Ba_{0.59}K_{0.41}Fe_2As_2$ the electron density of states is comprised of contributions from two electron bands with superconducting-state energy gaps that differ by a factor 3.8, with 77% coming from the band with the larger gap. The variation of the specific heat with magnetic field is consistent with extended $s$–wave pairing, one of the theoretical predictions. The relation between the densities of states and the energy gaps in the two bands is not consistent with a theoretical model based on interband interactions alone. Comparison of the normal-state density of states with band-structure calculations shows an extraordinarily large effective mass enhancement, for which there is no precedent in similar materials and no theoretical explanation.

# I. INTRODUCTION

There is considerable current interest in the possible existence of superconductors in which the mechanism of the electron pairing involves interactions other than the phonon-mediated electron-electron interaction of the BCS theory. The most likely candidates are "exotic" superconductors with unusual magnetic properties and a high critical temperature ($T_c$). Obvious similarities to other "simple" superconductors notwithstanding, the unusually high $T_c$ of $MgB_2$ attracted attention. As a consequence of the resulting research activity, the superconductivity of $MgB_2$ is perhaps better understood than any other. It is unusual in showing multiple energy gaps in the superconducting state (the first to be clearly identified as such) but it is fully accounted for by the BCS phonon-mediated interaction. The interest in other mechanisms has therefor been redirected to families of superconductors for which there are other reasons to believe that the superconductivity may involve different interactions. The doped Fe pnictides are the most recent to be discovered, and they are the subject of substantial research activity, both experimental and theoretical. The order parameter is an important key to understanding the mechanism, but in spite of the intense research effort the nature of the order parameter in the Fe pnictides is still not well understood. The conduction electron contribution to the specific heat ($C_e$) offers a useful approach to obtaining more information, and it has the advantage of being a bulk property, not sensitive to surface effects. In the superconducting and vortex states ($C_{es}$ and $C_{ev}$, respectively) its dependences on temperature ($T$) and magnetic field ($H$) are directly related to the number and magnitude of the energy gaps and the overall symmetry of the order parameter, including the existence of nodes. In the normal state ($C_{en}$) it is also the most important measure of the electron density of states (DOS), which is of critical importance in understanding the superconductivity. In the conventional analyses of the data this approach to obtaining information on the nature of the superconductivity has required an independent, necessarily approximate, determination of the lattice contribution to the specific heat ($C_{lat}$), which is subtracted from the total measured specific heat ($C$) to obtain $C_e$. However, at least for an optimally doped sample, the usual method of determining $C_{lat}$, measurement of $C$ in the normal state at temperatures up to $T_c$, is precluded by the high $T_c$ and high upper critical field ($H_{c2}$). In the 122 series of Fe-pnictide superconductors, doped $BaFe_2As_2$, a high-$T$ structural/magnetic transition, which is shifted to lower temperatures by the doping that produces the superconductivity, poses an additional obstacle to obtaining $C_{lat}$: In $BaFe_2As_2$ the transition is first-order and occurs near 140 K,[1] but in the K-doped superconductors, $Ba_{1-x}K_xFe_2As_2$, it is completely suppressed[2,3] at x ~ 0.35. Superconductivity occurs in the high-$T$ tetragonal phase for optimally doped (x ~ 0.4) and overdoped samples,[2-5] but in the low-$T$ orthorhombic phase for underdoped (x ~ 0.1 − 0.2) samples.[2,3,4]

Essentially all specific-heat data for high-$T_c$ Fe-pnictide superconductors have been analyzed by the same two-step procedure: In the first step, which is typical of that taken for any high-$T_c$ superconductor, an approximation for $C_{lat}$ was obtained in a fit to high-$T$ normal-state data, extrapolated to low temperatures, and subtracted from $C$ to obtain $C_{es}$. In the second step, the derived $C_{es}$ was analyzed with expressions based on the $\alpha$ model[6] as extended to two-gap superconductors[7] to derive the parameters characteristic of the electron bands. For five near-optimally hole-doped superconductors in the 122 series widely ranging values of the derived parameters have been reported. In some cases they have been interpreted as showing the presence of two energy gaps, but in others no evidence of the smaller gap was recognized. In addition, there are substantial inconsistencies in the results for the DOS. Here we suggest that errors associated with the different approximations made in the high-$T$ fit to obtain $C_{lat}$ and its extrapolation to low temperatures make a significant contribution to these ambiguities and inconsistencies: At the low temperatures at which evidence of the small gap is found, $C_{es}$ is much smaller than $C_{lat}$, and errors in the $C_{lat}$ obtained in the high-$T$ fit are magnified in the low-$T$ $C_{es}$, and carried through to the derived parameters.

In this paper we report the use of a new method to analyze specific-heat data for a high-$T_c$ superconductor that bypasses the need in the conventional analyses for a $C_{lat}$ that is subtracted from $C$ to obtain $C_{es}$. It eliminates the uncertainties and errors in $C_{es}$ produced by the approximations inherent in the determination of $C_{lat}$. It is based on comparisons of $\alpha$-model expressions for the electron contribution with the total measured specific heat, and gives the parameters characteristic of the electron bands directly. The parameters for a small-gap band are obtained from an analysis of the low-$T$ data. The parameters for a large-gap band are obtained from the discontinuities in $C$ and d$C$/d$T$ at $T_c$, after correcting for the contributions of the small-gap band. Here the analysis is applied to measurements on a near-optimally hole-doped Fe-pnictide superconductor in the 122 series, $Ba_{0.59}K_{0.41}Fe_2As_2$. A summary of the relations used to represent the different contributions to the specific heat and descriptions of the approximations used in obtaining $C_{lat}$ that would affect the derived $C_{es}$ are included in Sec. II. The sample and the measurements are described in Sec. III; the specific-heat results and the analysis of the data in Sec. IV. The results are discussed in Sec. V, compared with the results of conventional analyses of measurements on five other near optimally hole-doped 122 series superconductors in Sec. VI, and summarized in Sec. VII.

## II. CONTRIBUTIONS TO THE SPECIFIC HEAT; APPROXIMATIONS IN THE DETERMINATION OF THE LATTICE CONTRIBUTION

The normal-state electron contribution to $C$ is usually taken to be

$$C_{en} \equiv \gamma_n T_c, \tag{1}$$

where $\gamma_n$ is a temperature-independent constant (but see below) that is proportional to the DOS. If there are two bands $\gamma_n$ represents the sum of the two contributions. (When it is convenient to distinguish the specific-heat contributions or other properties of two bands, additional subscripts, 1 and 2, are used, e.g., $\gamma_n = \gamma_{n1} + \gamma_{n2}$, $C_{es} = C_{es1} + C_{es2}$, $\alpha_1$ and $\alpha_2$, etc.)

The superconducting-state electron contribution given by the BCS theory in the weak-coupling limit, has been tabulated by Mühlschlegel[8] in the form $C_{es}/\gamma_n T_c$ as a function of the reduced temperature, $t \equiv T/T_c$. Experimental results for 'strong-coupling materials" are inconsistent with this result, and they are also inconsistent with general limitations on the effects of strong coupling in the BCS theory.[6] This led to the formulation of the $\alpha$ model[6], a phenomenological extension of the BCS theory to include strong-coupling effects. In the $\alpha$ model the temperature dependence of the energy gap is taken to be that calculated[8] for the BCS theory in the weak-coupling limit, but the amplitude of the gap at $T = 0$, $\Delta(0)$, is an adjustable parameter represented by $\alpha \equiv \Delta(0)/k_B T_c$ that provides an empirical measure of the strength of the coupling. In the weak-coupling limit of the BCS theory $\alpha = 1.764 \equiv \alpha_{BCS}$. Early applications were focused on superconductors that showed other evidence of strong coupling, which gave values of $\alpha$ greater than $\alpha_{BCS}$, but for some superconductors the thermodynamic properties were represented by values of $\alpha$ less than $\alpha_{BCS}$, and recently this has been interpreted in terms of weak coupling. For $MgB_2$ at the lowest temperatures $C_{es}$ shows a large excess over that given by the BCS theory. It was recognized that this could be represented by the $\alpha$ model with $\alpha$ *much* less than $\alpha_{BCS}$,[7] which, however, would not be consistent with $C_{es}$ near $T_c$ (see Fig. 2 of Ref. 7). This suggested the extension of the $\alpha$ model to a two-band two-gap superconductor in which $C_e$ is taken to be the sum of two independent additive contributions, even though the equality of $T_c$ in the two bands requires some interband coupling.[7] The $\alpha$-model fits represent $C_{es}$ for $MgB_2$ to within the experimental accuracy, giving[7] $\alpha$ values of 2.2 and 0.6, which are consistent with detailed theoretical calculations[9] that show both strong and weak coupling. Currently, essentially all specific heat measurements on high-$T_c$ Fe pnictide superconductors are compared with a model of this kind, in which $C_{es}$ is represented by the sum of contributions with $\alpha$-model $T$ dependences and different values of $\alpha$.

The vortex-state electron contribution of a superconductor with an isotropic gap includes two terms:

$$C_{ev}(H) = C_{evs}(H) + \gamma_v(H)T. \qquad (2)$$

The first term, $C_{evs}(H)$, which is associated with the residual superconducting condensate, is the in-field counterpart of $C_{es}$ in zero field. It decreases in magnitude with increasing $H$ but the details of its $H$ and $T$ dependences are not theoretically established. The other term, $\gamma_v(H)T$, is associated with the vortex cores[10]. Its coefficient varies from $\gamma_v(0) = 0$ to $\gamma_v(H_{c2}) = \gamma_n$, with a variation that is, at least for a single-band superconductor, linear in $H$. In most samples of superconducting materials there is a "residual" DOS that produces a normal-state-like contribution to $C$ even in zero field. This appears as a non-zero value of $\gamma_v(0)$, $\gamma_r \equiv \gamma_v(0) \neq 0$, and is generally attributed to non-superconducting regions of the same material.

In the low-$T$ limit the lattice contribution can be represented by

$$C_{lat} = B_3 T^3 + B_5 T^5 + B_7 T^7 + \text{- - - -},\tag{3}$$

where $B_3$ is the coefficient of the $T^3$ term of the Debye theory,

$$B_3 = (12/5)\pi^4 R/(\theta_D)^3,\tag{4}$$

and $\theta_D$ is the Debye characteristic temperature. The higher-order terms represent the effects of phonon dispersion, and they may also serve as an approximation for the low-$T$ contributions of low-frequency optical modes if the lattice has a basis. However, Eq. (3) is often used in an interval of temperature at higher temperatures, in which case it is just a convenient fitting expression with no physical meaning. In particular, coefficients obtained in the high-$T$ fits cannot be expected to give a valid expression for $C_{lat}$ at lower temperatures. Combinations of Debye and Einstein functions are also used to represent $C_{lat}$ at higher temperatures, where they are physically more reasonable fitting expressions, but the fits are relatively insensitive to the values of the fitting parameters, and the parameters derived, like those derived from high-$T$ fits with Eq. (3), should not be expected to give $C_{lat}$ accurately at lower temperatures.

To obtain the $H$ and $T$ dependences of $C_e(H)$ for $T \leq T_c$ in a conventional analysis, it is necessary to have an expression for $C_{lat}$ that is valid in the same temperature interval. For the Fe pnictides two distinctly different methods for obtaining an approximation for $C_{lat}$ have been used. In one, the first step is to obtain $C_{lat}$ for $T \leq T_c$ for a comparison material for which the normal-state specific heat is known. The comparison materials that have been used include the undoped non-superconducting parent compound, an overdoped non-superconducting sample, and a material with a different dopant that suppresses both the superconductivity and the high-$T$ structural/magnetic transition. In some cases adjustments to $C_{lat}$ of the comparison material for the differences in stoichiometry or structure are made, but they are necessarily rough approximations. Furthermore, the effect on $C_{lat}$ of the substantial differences in the DOS are quite generally ignored. The other method is to obtain $C_{lat}$ for the sample itself by fitting the normal-state data for $T \geq T_c$ with $C = \gamma_n T + C_{lat}$, and extrapolating the resulting $C_{lat}$ to $T < T_c$ to determine $C_{es}$. In addition to the fact that an expression obtained for $C_{lat}$ in a high-$T$ interval cannot be expected to be accurate at low temperatures, there are other reasons for doubting the validity of the derived $C_{lat}$ (and also $\gamma_n$, if it is derived in the fit): Since $C$ is measured at constant pressure it includes a contribution to $C_{lat}$ associated with the anharmonicity of the lattice vibrations that can also be approximately $T$ proportional.[11] For samples of $Ba_{1-x}K_xFe_2As_2$ this contribution has been estimated[12] to increase rapidly from zero at $T = 0$ to $\sim 600$ mJ K$^{-1}$ mol$^{-1}$ at 100 K, to increase less rapidly at higher temperatures, and to become more nearly $T$ proportional above 150 K, with a coefficient $\sim 12$ mJ K$^{-2}$ mol$^{-1}$. Furthermore, the phonon enhancement that contributes to $\gamma_n$, and therefore $\gamma_n$ itself, is expected to be $T$ dependent (see, e.g., Refs. 13 and 14). The complicated temperature dependence of $C_{lat}$, including the anharmonic contribution, prevents the identification of this effect in specific-heat measurements, but there is compelling evidence for its reality in cyclotron resonance experiments.[15] There is no basis for estimating its magnitude in these materials, but it could be substantial. The difficulties associated with obtaining an independent approximation for $C_{lat}$, ensure substantial uncertainty in any $C_{es}$ obtained in the conventional analyses.

The determination of $C_{lat}$ is the major obstacle to obtaining $C_e$ from experimental data, but in most samples there are paramagnetic centers that also make a significant contribution to $C$, which is best represented by an $H$-dependent approximation to a Schottky function, $C_{Sch}(H)$. With this contribution, the total specific heat in a field $H$ is

$$C(H) = C_{lat} + C_e(H) + \text{m}C_{Sch}(H), \tag{5}$$

where m is the molar concentration of paramagnetic centers. For $0 \leq H < H_{c1}$, where $H_{c1}$ is the lower critical field (and omitting the possible $\gamma_t T$ contribution) $C_e(H) = C_{es}$; for $H_{c1} \leq H < H_{c2}$, $C_e(H) = C_{evs}(H) + \gamma_v(H)T$; for $H \geq H_{c2}$, $C_e(H) = \gamma_n T$.

The requirement of entropy conservation, the equality of the conduction-electron entropies in the normal and superconducting states at $T_c$, is frequently invoked, either as a constraint in a fitting procedure used to obtain $C_{lat}$ or as a test of the validity of a derived $C_{lat}$. In zero field it takes the form

$$\int (C_{es}/T)\mathrm{d}T = \gamma_n T_c, \tag{6}$$

where $C_{es} = C - C_{lat}$, $- \text{m}C_{Sch}$ and the integration extends from $T = 0$ to $T = T_c$. In the special case of a $C_{lat}$ that is determined in a high-temperature fit to normal-state data and then extrapolated to low temperatures, imposition of the entropy-conservation constraint can reduce gross errors in the derived $C_{lat}$ (see Sec. VI). More generally however, its effect is limited by the small fraction of the entropy at $T_c$ that is electron entropy, e.g., ~13% in the results reported here. At best, even if an accurate value of $\gamma_n$ is known independently, satisfaction of Eq. (6) shows only that $C_{es}$ gives the correct *entropy* at $T_c$, i.e., that it is only a $T^{-1}$-weighted *average* of $C_{es}$ that is correct. This leaves room for $T$-dependent errors that are comparable in magnitude to small contributions to $C_{es}$ that have been attributed to small-gap bands in temperature intervals near or below $T_c/2$. Furthermore, in many cases the validity of the value of $\gamma_n$ used in Eq. (6) is not obvious, and in some cases its origin is not clearly specified.

## III. SAMPLES AND MEASUREMENTS

Single crystals of $Ba_{0.59}K_{0.41}Fe_2As_2$ were synthesized by a self-flux method[16]. The stoichiometry was checked by inductively coupled plasma and electron microprobe wavelength-dispersive X-ray spectroscopy. As shown in the inset to Fig. 1, ac-susceptibility measurements for $\mu_0 H = 10^{-4}$ T, made with the ACMS option of the Physical Property Measurement System (PPMS) of Quantum Design, showed a sharp step-like transition in $\chi'$ with a width of under 2 K, and full superconductivity. There was no indication of an anomaly in the specific heat in the vicinity of 70 K (see Fig. 1) that would indicate the presence of FeAs, which is a common impurity in samples of these materials (see, e.g., Refs. 3 and 12). (The "glitch" near 80 K marks the transition between different specific-heat runs.) For a sample of $Ba_{0.55}K_{0.45}Fe_2As_2$ that *did* show the FeAs anomaly at 70 K, the discontinuity in $C$ at $T_c$ was only 2/3 of that reported here. Since the magnitude of the discontinuity at $T_c$ plays an important role in the interpretation of specific-heat data, a reduced magnitude of the discontinuity associated with the presence of FeAs would have significant consequences. In addition to the susceptibility measurements and the absence of a detectable level of FeAs, the absence of a residual DOS ($\gamma_r = 0$), the low concentration of paramagnetic centers, and the relatively sharp anomaly in the specific heat at $T_c$ (see Sec. IV) attest the high quality of the sample.

The specific heat of a 10.3-mg, plate-like single crystal was measured in the PPMS from 2 to 300 K in zero field. Below 50 K measurements were also made in 9 fields applied perpendicular to the *ab* plane to a maximum $\mu_0 H = 14$ T. A different set of measurements on the same sample was reported in an earlier paper[17]. The measurements reported here were made after the sample had aged for a longer time at room temperature, and the results are slightly different, but the main feature of the anomaly at $T_c$, the discontinuity in $C$, is essentially the same. To obtain more accurate data than those reported in Ref. 17, the specific heat of the addenda and the sample were measured at the same temperatures, and the platform thermometer was calibrated in each of the fields in which the specific heat was measured.

## IV. SPECIFIC-HEAT RESULTS AND ANALYSIS

The specific heat results for $H = 0$ are shown for 2 to 300K in Fig. 1, and on expanded scales to lower temperatures in Fig. 2. The discontinuity in $C$ at 36.9 K (see also Fig. 3) marks the transition to the superconducting state. The solid sloping lines in Fig. 3, which represent the ideally sharp mean-field transition in zero field, are the results of somewhat arbitrary, but typical, straight-line fits to the data just outside the region of curvature associated with the broadening of the transition by sample inhomogeneity and fluctuation effects. Their extrapolations to $T_c$, together with the entropy-conserving dash-dot vertical line, determine $T_c$ as 36.9 K. Since $C_{lat}$ is continuous at $T_c$, the solid lines give the discontinuity in $C_e$, $\Delta C_e(T_c)/T_c = 157.5$ mJ K$^{-2}$ mol$^{-1}$;

with some mathematical manipulation, $dC/dT = T d(C/T)/dT + C/T$, they also give the discontinuity in $dC_e/dT$, $\Delta(dC_e/dT)|_{Tc} = 1183$ mJ K$^{-2}$ mol$^{-1}$. In comparison with other measurements on similar materials the transition is relatively sharp and the discontinuities are relatively large.

The first step in the analysis is to obtain approximate, preliminary values of $\gamma_n$ and $\alpha$ from the data in the vicinity of $T_c$ using $\alpha$-model expressions for a single gap. The $\alpha$ model gives the discontinuities in $C_e$ and $dC_e/dT$ in terms of the parameters $\gamma_n$ and $\alpha$. Conversely, it can be used to obtain $\gamma_n$ and $\alpha$ from the experimental values of the two discontinuities. For any value of $\alpha$ it gives $C_{es}$ as a function of $t = T/T_c$,

$$C_{es}(t)/\gamma_n \, T_c \equiv \mathrm{f}_\alpha(t). \tag{7}$$

Since $C_{en} = \gamma_n T$, $C_{en}(T_c)/T_c = \gamma_n$, and the discontinuity in $C_e$ at $T_c$ is

$$\Delta C_e(T_c)/T_c = C_{es}(T_c)/T_c - C_{en}(T_c)/T_c = \gamma_n[\mathrm{f}_\alpha(1) - 1]. \tag{8}$$

Since $(dC_{es}/dt)/\gamma_n T_c = d\mathrm{f}_\alpha(t)/dt \equiv \mathrm{f}_\alpha'(t) = (dC_{es}/dT)/\gamma_n$, and $(dC_{en}/dT)|_{Tc} = \gamma_n$, the discontinuity in $dC_e/dT$ *is*

$$\Delta(dC_e/dT)|_{Tc} = (dC_{es}/dT)|_{Tc} - (dC_{en}/dT)|_{Tc} = \gamma_n[\mathrm{f}_\alpha'(1) - 1]. \tag{9}$$

If $\gamma_n$ is known independently, either Eq. (8) or Eq. (9) would give the value of $\alpha$, and both of these equations have been used in that way. However, taken together, the two equations can be used to obtain the value of $\alpha$ independently of $\gamma_n$: The ratios of the left and right hand sides of Eqs. (8) and (9) give

$$T_c\Delta(dC_e/dT)|_{Tc}/\Delta C_e(T_c) = [\mathrm{f}_\alpha'(1) - 1]/[\mathrm{f}_\alpha(1) - 1], \tag{10}$$

which determines the value of $\alpha$ as that for which the function of $\alpha$ on the right-hand side agrees with the experimental quantity on the left. With the value of $\alpha$ determined by Eq. (10), Eq. (8) or Eq. (9) can be used to obtain $\gamma_n$. In the present case the result is $\gamma_n = 32.2$ mJ K$^{-2}$ mol$^{-1}$, and $\alpha = 3.27$. These would be the correct values if there were only a single band, but if there is also a small-gap band the discontinuities would have to be corrected for its contributions and the parameters of the large-gap band recalculated.

The test for the existence of a small-gap band was based on a search for its contribution to $C(H)$ in a "global" fit with Eq. (5) to the data for all $H$ and for $T \leq 12$ K. The details of the final fitting expression were based on the results of trials of a number of different fitting expressions and different temperature intervals for the fits. The results of some of these preliminary fits are described, together with other evidence of the validity of the fit, in the final paragraph of this section. The final fitting expression made allowance for four contributions to $C(H)$: the contribution of the lattice, represented by three terms of Eq. (3); the contribution of the vortex-cores, represented by $\gamma_v(H)T$; the contribution of the superconducting condensate, $C_{es}$ for $H = 0$, and $C_{evs}$ for $H \neq 0$; a contribution of paramagnetic centers, represented by a two-level Schottky function (see below) with an $H$-dependent characteristic temperature, $\theta_{Sch}(H) = \theta_{Sch}(0)(1 + \beta H^2)^{1/2}$. Inclusion of the paramagnetic-center contribution was suggested by the deviations from linearity in the plot of $C/T$ vs $T^2$ (see Fig. 4) which are typical indications of the presence of a low concentration of paramagnetic centers.

For $T \leq 12$ K the component of $C_{es}$ associated with a large-gap band with $\alpha \sim 3$ and $\gamma_n \sim 30$ mJ K$^{-2}$ mol$^{-1}$ would be negligible, and it is only the component associated with a small-gap band that needs to be considered. As given by Eq. (7), that component would be $C_{es2}(T) = \gamma_{n2}T_c f_\alpha(T)$. For this interval of temperature and the values of $\alpha$ that turn out to be of interest ($\sim 1$) it can be represented by the exponential of a three-term polynomial in $T^{-1}$, $-X_\alpha(T)$, giving

$$C_{es2}(T) = \gamma_{n2}T_c f_\alpha(T) = \gamma_{n2}T_c \exp[-X_\alpha(T)], \tag{11}$$

with the three coefficients in $X_\alpha(T)$ determined by the value of $\alpha$. Generalizing that expression to extend its validity to the in-field data, i.e., to include $C_{evs2}(H)$ for $H \neq 0$, requires allowing for its $H$ dependence in the vortex state. There is little theoretical guidance for such a generalization, and it was made empirically. Experimental results on other superconductors suggest two changes: the replacement of the pre-exponential coefficient, $\gamma_{n2}T_c$, with an $H$-dependent coefficient, $a(H)$, to allow for the reduction in the magnitude of the residual superconducting condensate contribution that is complementary to the development of the vortex core contribution, and the inclusion of an $H$-dependent factor, $b(H)$, in the exponent to allow for the effective reduction of the gap by the excitation of quasiparticles within the gap. With these changes, the component of $C_{evs}(H)$ and $C_{es}$ associated with the small-gap band is $a(H)\exp[-b(H) X_\alpha(T)]$ and the fitting expression becomes

$$C(H) = C_{lat}(T) + \gamma_v(H)T + a(H)\exp[-b(H)X_\alpha(T)] + mC_{Sch}(H, T). \tag{12}$$

Fitting the data for all $H$ simultaneously more than doubles the ratio of number of points in the fit to number of adjustable parameters, and gives more reliable values of the parameters. It is also desirable for the information it gives about the $H$ dependences of the contributions, e.g., $\gamma_v(H)$. However, with a high density of more precise, more accurate data, which could be obtained in other apparatus, it might be possible to get a good fit to the zero-field data alone, without resorting to the $H$ dependence introduced empirically in the third term.

A fit has to be made for a specified value of $\alpha$, which determines the values of the three fixed parameters in $X_\alpha(T)$, and the derived values of the adjustable parameters depend on the value for which the fit was made. The third term in Eq. (12) represents the component of the contribution of the superconducting condensate coming from the small-gap band in *all* fields. In zero field its $T$ dependence is not correct for *any* value of $\alpha$ unless $b(0) = 1$. This provides the criterion for recognizing the correct value of $\alpha$, i.e., that for which the fit gives $b(0) = 1$. For the same reason, that fit gives $\gamma_{n2}$, as $\gamma_{n2} = a(0)/T_c$. The strong dependence of $b(0)$ on $\alpha$ — e.g., $b(0) = 1.109, 0.946$, and $0.822$ for $\alpha = 0.8, 0.9$, and $1.0$ — ensures a precise determination of the value of $\alpha$. The result $b(0) = 1$ was obtained for $\alpha = 0.86$, and for that fit $a(0) = 337 \pm 17$ mJ K$^{-1}$ mol$^{-1}$. These results show the existence of a small-gap band characterized by the parameters $\alpha_2 = 0.86$ and $\gamma_{n2} = 9.1 \pm 0.5$ mJ K$^{-2}$ mol$^{-1}$. Because of the small value of $\gamma_{n2}$, and particularly the small value of $\alpha_2$, that band makes only small contributions to the discontinuities at $T_c$: $\Delta C_{e2}(T_c)/T_c = 3.1$ mJ K$^{-2}$ mol$^{-1}$; $\Delta[(\mathrm{d}C_{e2}/\mathrm{d}T)|_{Tc}] = 3.5$ mJ K$^{-2}$ mol$^{-1}$. Correcting the measured discontinuities for these contributions gives $\Delta C_{e1}(T_c)/T_c = 154.4$ mJ K$^{-2}$ mol$^{-1}$ and $\Delta[(\mathrm{d}C_{e1}/\mathrm{d}T)|_{Tc}] = 1180$ mJ K$^{-2}$ mol$^{-1}$, which in turn give $\alpha_1 = 3.30$, $\gamma_{n1} = 31.0$ mJ K$^{-2}$ mol$^{-1}$, and a total $\gamma_n = 40.1$ mJ K$^{-2}$ mol$^{-1}$. Parameters that characterize the two bands are listed in Table I.

The other contributions to $C(H)$ obtained in the fit are plausible and consistent with the behavior known in other superconductors. The $H$-independent parameters obtained in the fit are: m = $1.29 \pm 0.15$ x $10^{-3}$ mol mol$^{-1}$; $\theta_{\mathrm{Sch}}(0) = 7.32 \pm 0.41$ K; $\beta = 3.30 \pm 1.17$ x $10^{-2}$ T$^{-1/2}$; B$_3$ = $0.602 \pm 0.022$ mJ K$^{-4}$ mol$^{-1}$; B$_5$ = $7.23 \pm 2.1$ x $10^{-4}$ mJ K$^{-6}$ mol$^{-1}$; B$_7$ = - $6.1 \pm 7.0$ x $10^{-7}$ mJ K$^{-8}$ mol$^{-1}$. The $H$-dependent parameters are given in Table II, and $\gamma_v(H)$ is displayed graphically in Fig. 5. The evolution with increasing $H$ of each of the three $H$-dependent contributions to $C(H)$ is illustrated in Fig. 6 for $\mu_0 H = 0, 6$, and 14 T, with the $H$-independent $C_{lat}$ included for comparison. The $H$ dependences of the overall magnitude of the contribution of the superconducting condensate and of the energy gap that were introduced empirically, the factors $a(H)$ and $b(H)$, give a satisfactory representation of the experimental data. Furthermore, and as expected, the results of the fit are consistent with the behavior seen in measurements on other superconductors: The contribution of the superconducting condensate decreases with increasing $H$, as shown by both the values of $a(H)$ and the plots of $C_{es2}$ and $C_{evs2}$ in Fig. 6. The $T$ and $H$ dependences of $C_{es2}$ and $C_{evs2}$ are plausible, and the exponential downturns at low temperatures occur at temperatures consistent with the values of $b(H)$ in showing the expected decrease in the effective gap with increasing $H$.

The identification of a band with a small energy gap requires the accurate determination of $C_{es}$ at the low temperatures at which it would make a significant contribution to $C(0)$, typically $T \leq 15$ K for the small gaps that have been reported in these materials. The problem, *for any analysis*, is to separate the small $C_{es}$ — a maximum of only 12% of $C(0)$ near 9 K in our results — from the much greater $C_{lat}$. Our analysis depends on the validity of the fitting expression, Eq. (12), and the fit to $C(H)$. Unlike the conventional analyses, it does not involve a pre-determined quantitative expression for $C_{lat}$. However, it does require inclusion in the fitting expression of a contribution that has the $T$ dependence generally expected for $C_{lat}$ in the low-$T$ limit (see Sec. II), and this was taken to be the sum of $T^3$, $T^5$, and $T^7$ terms, the first three terms in Eq. (3), with adjustable coefficients. The numerical values of the coefficients are a necessary byproduct of the fit. They determine $C_{lat}$ in this limited temperature interval but they do not affect the derived $C_{es2}$. Eight different preliminary fits, with or without the $T^7$ term, to either 10 or 12 K, and for $\alpha$ either 0.8 or 0.9, gave the same amplitude of the contribution of the small-gap band to within ± 5%, and to within ± 2.5% for each group of four for which $\alpha$ was the same. These fits suggested that the $T^7$ term would make only a marginal contribution, but it was included in the final fitting expression to give maximum flexibility. The paramagnetic-center contribution presented the major problem with the fitting expression. The two-level Schottky anomaly in the final fitting expression is clearly too narrow in temperature, but broader anomalies that were tried — two-level Schottky anomalies with different degeneracies of the levels, two-level Schottky anomalies with Gaussian or Lorentzian broadening, and a three-level Schottky anomaly — made no significant improvement in the fit and did not suggest an alternative. There are 36 adjustable parameters in the fitting expression, including the 10 that allow for the $H$ dependence of $\gamma_v(H)$ and the 20 that model the expected $H$ dependences of the superconducting-condensate contribution, but there are 320 data points in the fit, an adequate excess over the number of parameters. The fit was made using a non-linear least-squares procedure in the Matlab computational language, and carried to the smallest convergence tolerance allowed in the Matlab program. To ensure that the fitting process converged to the best possible result (an absolute minimum of the reduced $\chi^2$) a number of fits were made with different initial values of the parameters and different iteration step sizes. The fractional deviations in the final fit are up to ± 3% at 2 K, where the Schottky contribution to C(H) is significant, but they are within ± 1% and ± 0.25%, respectively, at 6 and 12 K, the limits of the interval that is important for determining $C_{es2}$. The Schottky contribution is relatively small, and most significant at the lowest temperatures in low fields (see Fig. 6). Its small size accounts for the relatively large uncertainties in the parameters m, $\theta_{Sch}(0)$ and $\beta$. It is only the sharp drop off on the high-$T$ side, which is not sensitive to the details of the fitting expression, that is relevant to separating the four contributions to $C(H)$. For that reason, and because the Schottky anomaly is not of any interest in itself, the inadequacy of the fitting expression in representing it accurately not important. With that allowance for the Schottky contribution, the $T$ dependences of the four contributions are all well defined and substantially different. This is of considerable importance in connection with the validity of their separation, which is also supported by the relatively small uncertainties in the relevant parameters. The validity of the result for $C_{es2}$ is also *directly* supported by the strong dependence of $b(0)$ on $\alpha$ (see above) which is persuasive evidence of the existence of a term in $C(0)$ with a $T$ dependence corresponding to the contribution of a small-gap band with a value of $\alpha$ within the range of the fits. The difficulties in determining $C_{es2}$ notwithstanding, the evidence for a small-gap band characterized by $\alpha_2 = 0.86$ and $\gamma_{n2} = 9.1 \pm 0.5$ mJ K$^{-2}$ mol$^{-1}$ is reasonably strong.

## V. DISCUSSION

The major result of the analysis is the identification of two electron bands that contribute to the DOS, and have substantially different energy gaps in the superconducting state. As measured by the coefficients of the electron contributions to the specific heat, the total DOS, $\gamma_n = 40.1$ mJ $K^{-2}$ $mol^{-1}$, is comprised of a 77% contribution, $\gamma_{n1} = 31.0$ mJ $K^{-2}$ $mol^{-1}$, from the band with the larger gap, $\Delta_1(0) = 10.49$ meV, and a 23% contribution, $\gamma_{n2} = 9.1$ mJ $K^{-2}$ $mol^{-1}$, from the band with the smaller gap, $\Delta_2(0) = 2.73$ meV. The results for $C_{es}$ and its two components are shown graphically in Fig. 7, with the result of the BCS theory in the weak coupling limit and for the same $\gamma_n$, included for comparison. Although circumvention of the need for an independent determination of $C_{lat}$ is an important feature of our analysis, $C_{lat}$, and its relation to $C(0)$, is of some interest for comparison with the results of other measurements. The 2 – 12 K fit with Eq. (12) gives $C_{lat}$ for that temperature interval. At higher temperatures the apparent $C_{lat}$ can be obtained by subtracting $C_{es}$ or $C_{en}$ from $C(0)$, and for that purpose the actual $C(0)$ data in the immediate vicinity of $T_c$ were replaced by the straight lines in Fig. 3 that represent the idealized sharp transition. The results for $C_{lat}$ to 40 K, the limit of the straight-line construction in Fig. 3, are represented by the solid lines in Fig. 2. The small difference between $C(0)$ and $C_{lat}$ for $T \leq 20$ K emphasizes the sensitivity to errors in $C_{lat}$ of a $C_{es2}$ derived from that difference.

Several other techniques give values of the energy gaps that can be compared with those derived from the specific-heat data. Quite generally, the results obtained by these techniques suggest that there are two gaps with substantially different magnitudes in the Fe–pnictide superconductors (see, e.g., Ref. 18). Here we focus on those obtained from angle-resolved photoemission spectroscopy (ARPES) measurements on $Ba_{1-x}K_xFe_2As_2$, which are the most extensive and detailed of the other measurements. The comparison is best made on the basis of the values of $\Delta_1(0)$ and $\Delta_2(0)$, which are given directly by the ARPES results, and are independent of $T_c$. In the following. $\Delta_1(0)$ and $\Delta_2(0)$ are used for the larger and smaller gaps, respectively, regardless of the notation used in the other publications. As derived from the specific-heat data, these quantities are averages in the sense that small differences between different sheets of the Fermi surface and anisotropies on a single sheet are not resolved. ARPES measurements give more detailed information but the results are often summarized by two averages over narrow ranges of gap magnitude. For a sample with $T_c = 32$ K, Evtushinsky et al.[19] report $\Delta_1(0) = 9.2 \pm 1$ meV for an inner hole-like barrel at the $\Gamma$ point, and smaller gaps on all other elements of the Fermi surface. However, the feature that showed the opening of the larger gap was not observed for the smaller gaps, and they conclude only that $\Delta_2(0) < 4$ meV. For a sample with $T_c = 37$ K and x = 0.4, Ding et al.[20], report $\Delta_1(0) \sim 12.5$ meV for the inner $\Gamma$ barrel and $\Delta_2(0) \sim 5.5$ meV for the outer $\Gamma$ barrel, but the unusual temperature dependence of the gaps leaves some doubt about the extrapolation to 0 K. For a sample with $T_c = 35$ K and x = 0.4, Zhao et al.[21] report anisotropic gaps, $\Delta_1(0) = 10 - 12 \pm 1.5$ meV for the inner $\Gamma$ barrel, and $\Delta_2(0) = 7 - 8 \pm 1.5$ meV for the outer $\Gamma$ barrel. The two Fermi surface spots near the M point are gapped below $T_c$ but the gaps persist above $T_c$. For a sample with x = 0.45, but unspecified $T_c$, Liu et al.[22] report measurements on samples that "display bulk superconductivity" but the superconducting gaps are not detected in measurements at 12 K. Our value of $\Delta_1(0)$ falls well within the range of those obtained from ARPES results, but, while the value of $\Delta_2(0)$ is consistent with that obtained by Evtushinsky et al.[19], it is substantially lower than the other two ARPES values. Although our value was obtained from a small feature in the low-temperature specific heat, the sensitivity of the fits to the value of $\alpha_2$, which determines $\Delta_2(0)$, argues against such an error in $\Delta_2(0)$. Comparably small values of $\Delta_2(0)$, as measured by $\alpha_2$, have been reported in electron-doped $BaFe_2As_2 - \alpha_2 = 0.95$ in Ref. 23 and $\alpha_2 = 0.957$ in Ref. 24 – but, given the differences between the electron- and hole-doped compounds, the implications of this similarity in the values of $\alpha_2$ are not clear.

The $H$ dependence of $\gamma_v(H)$ gives information about the symmetry of the order parameter, most directly on the existence of nodes. For "conventional" s-wave superconductors with an isotropic gap, there is a normal-state-like electron contribution to the specific heat associated with the vortex cores[11], the $H$-proportional $\gamma_v(H)$ term described in Sec. II. For a d-wave superconductor Volovik predicted an $H^{1/2}$ dependence associated with extended quasiparticle states near line nodes.[25] This effect was first observed by Moler et al.[26] in a cuprate superconductor. It has been suggested that this $H^{1/2}$ dependence is modified to $H\ln H$ at low fields in a dirty superconductor.[27] Modifications of the $H$-proportional dependence in the case of an isotropic gap, negative curvature in high fields, have also been suggested[28]. The $H$ dependence of $\gamma_v(H)$ is compared with $H$ and $H^{1/2}$ dependences in Fig. 5. Overall, $\gamma_v(H)$ is better represented by the solid straight line, which has a slope 0.75 mJ $K^{-2}$ $mol^{-1}$ $T^{-1}$, than the dashed curve for $H^{1/2}$. (The $H\ln H$ dependence suggested for a dirty d-wave superconductor[27] would not give a better fit.) For this reason, and particularly because the low-field data suggest a finite limiting slope, these results are more consistent with an isotropic gap than with the low-energy excitations associated with nodes. Two other measurements of $\gamma_v(H)$, to 9 T, on similar hole-doped $BaFe_2As_2$ samples have been interpreted in the same way: Apart from non-zero values of $\gamma_v(0)$, an approximately $H$-proportional dependence[29] with a slope 0.63 mJ $K^{-2}$ $mol^{-1}$ $T^{-1}$, and a more precisely determined $H$-proportional dependence[30] with a slope 0.60 mJ $K^{-2}$ $mol^{-1}$ $T^{-1}$ have been reported. There is no obvious explanation for the curvature in $\gamma_v(H)$ in Fig. 5. The curvature predicted for an isotropic gap[28] seems to be significant only at higher fields. However, for $MgB_2$ there is a relatively sharp bend in $\gamma_v(H)$ vs $H$ that is associated with different values of $H_{c2}$ for the two bands[31], and perhaps an effect of that kind, but with a smaller difference in the values of $H_{c2}$, could be at work here.

Band-structure calculations[32] for $Ba_{1-x}K_xFe_2As_2$ using the local-density approximation (LDA), the virtual-crystal model, and allowing the positions of the As atoms to relax according to the LDA energy minimization criterion, show a very weak dependence of the DOS on doping. For the undoped $BaFe_2As_2$ the "bare" band-structure DOS is $N(E_F) = 3.06$ states $eV^{-1}$ f.u.$^{-1}$; for the x = 0.4 hole-doped material $N(E_F) \sim 3.12$ states $eV^{-1}$ f.u.$^{-1}$. However, the rigid-band calculation[32], which gave essentially the same result for x = 0, gave $N(E_F) \sim 4.38$ states $eV^{-1}$ f.u.$^{-1}$ for x = 0.4.[32] Another calculation[33] gave $N(E_F) = 4.553$ states $eV^{-1}$ f.u.$^{-1}$ for $BaFe_2As_2$, and, using a supercell model, $N(E_F) = 5.526$ states $eV^{-1}$ f.u.$^{-1}$ for x = 0.5. The increase in $N(E_F)$ for x = 0.5 in that calculation was ascribed to the use of the fixed experimental As position for the undoped compound.[32] For comparison with experimental quantities, we take, somewhat arbitrarily, the value $N(E_F) = 3.12$ states $eV^{-1}$ f.u.$^{-1}$ from Ref. 32. The corresponding contribution to the electron specific heat, the "bare" band-structure DOS, represented as a component of $\gamma_n$, is $\gamma_0 = 7.35$ mJ $K^{-2}$ $mol^{-1}$. The experimental value of $\gamma_n$, 40.1 mJ $K^{-2}$ $mol^{-1}$, then suggests an effective mass renormalization that would be unusually strong for a simple metal, for which the mass renormalization is produced by the electron-phonon interaction represented by the electron-phonon coupling parameter $(\lambda)$ and $\gamma_n = (1 + \lambda)\gamma_0$. The value of $\lambda$ would be 4.5, a factor 10 or so higher than the values commonly attributed to the electron-phonon interaction. The theoretical value of $N(E_F)$ chosen for the comparison was among the lowest, but the experimental value of $\gamma_n$ was also among the lowest (see Sec. VI), and any of the possible comparisons would still give an extraordinarily high value of $\lambda$. Although the mass renormalization for F-doped LaOFeAs, in the 1111 series of Fe pnictide superconductors, is not as strong as that found here for a member of the 122 series, it is strong enough to have attracted attention and it has motivated several calculations of the electron-phonon interaction. In one calculation[34], the electron-phonon $\lambda$ was found to be ~ 0.2, and in another[35] 0.21. In both cases it was concluded that these numbers are too small to explain the apparent mass renormalization, and that the electron-phonon interaction is also too weak to account for the observed $T_c$. There are differences between the 1111 and 122 series, but, since the superconductivity occurs in the FeAs layers in both, it is reasonable to assume that these conclusions, with some allowance for differences in the numbers, would apply to $Ba_{1-x}K_xFe_2As_2$. It therefore seems likely that calculation of the electron-phonon interaction for $Ba_{0.59}K_{0.41}Fe_2As_2$ would not account for the observed mass enhancement.

The electron-phonon interaction accounts for both the normal-state mass renormalization and the superconducting-state electron pairing in "conventional" superconductors. The fact that it doesn't account for either in the Fe pnictides raises the question as to whether there is another interaction that contributes to both. Interaction with spin fluctuations, which can support spin-singlet superconductivity only if there is a sign-changing order parameter, has been suggested as the mechanism for the electron pairing.[34] It was further suggested that the pairing would be "extended" s wave, designated $s_\pm$, in which isotropic order parameters on different sheets of the Fermi surface have opposite signs.[34] The approximate linearity of $\gamma_v(H)$ in $H$ (see Fig. 5) supports the argument in Ref. 34 that the $s_\pm$ pairing is more likely than d-wave, which could also satisfy the requirement of a sign-changing order parameter, but which would have nodes in the energy gaps. With respect to the mass renormalization, it is suggested in Ref. 36, which includes a general comparison of the superconductivity in the 1111 and 122 series, that while spin fluctuations might produce the strong mass enhancement in the 1111 series they might not produce the stronger effect in the 122 series. However, there seem to be no quantitative calculations. The specific-heat results emphasize the importance of theoretical consideration of magnetically mediated electron-electron interactions and their role in both mass enhancement and the occurrence of superconductivity. In connection with other theoretical predictions, we note that, in common with most other experimental work, the relations between energy gaps and the DOS that we report seem to be inconsistent with a theory of the superconductivity[37] based solely on interband interactions.

## VI. COMPARISON WITH THE RESULTS OF OTHER SPECIFIC-HEAT MEASUREMENTS

In this section we compare our results with those derived from five other specific-heat measurements on near-optimally hole-doped $BaFe_2As_2$. The focus is on the parameters that characterize the electron bands, which were obtained by conventional analyses based on a determination of $C_{lat}$ for the other measurements. The purpose is twofold: to consider the possibility that the approximations used for $C_{lat}$ produced significant uncertainties in the values of the parameters derived from the other measurements, and, with the results of those considerations in mind, to consider whether the other measurements suggest errors in the values that we report. The relevant features of the other measurements and the analyses of the data are summarized in the subsections A to E below, and those letters are used in the following to refer to both the subsections and the references. Particular attention is given to the values of $B_3$, the coefficient of the $T^3$ term in Eq. (3) for $C_{lat}$. This term is of major importance in the determination of $C_{es}$ in the vicinity of 10 - 15 K, where the contributions of small-gap bands have usually been reported. In that region it accounts for a significant fraction of $C_{lat}$, up to 90 % for the $Ba_{0.59}K_{0.41}Fe_2As_2$ sample described here. It can usually be obtained by analysis of the low-temperature data, but the effective value in the approximation for $C_{lat}$ that is used to obtain $C_{es}$ is often different. Since its role in the analysis can have an effect on the resulting value of $\gamma_n$, satisfaction of the entropy-conservation requirement is also noted. Our values of the parameters are compared with those derived from the other measurements in Table III. The subscripts 1 and 2 are used for the large- and small-gap bands, respectively, regardless of the notation used in the other publications.

### A. $Ba_{0.6}K_{0.4}Fe_2As_2$: Welp et al.[29];Mu et al.[38]

The specific heat of a sample with $T_c = 35.8$ K was measured from 2 to 50 K. The low-$T$ data gave $B_3$ and a $\gamma_v(H)$ that was approximately linear in $H$. Extrapolation of $\gamma_v(H)$ to $H_{c2}$, which was taken to be 100 T, gave $\gamma_n = 71$ mJ $K^{-2}$ $mol^{-1}$, which included a residual $\gamma_r = 7.7$ mJ $K^{-2}$ $mol^{-1}$. $C_{lat}$, represented by five terms in Eq (3) with $B_3$ fixed at the value derived in the low-$T$ fit, was obtained by fitting the normal-state data in the narrow interval above $T_c$ with $\gamma_n$ fixed at 71 mJ $K^{-2}$ $mol^{-1}$. Entropy conservation was "satisfied naturally". A single-gap $\alpha$-model fit to $C_{es}$, which gave $\alpha = 1.9$, represented the data below 13 K but did not account for a "hump" at higher temperatures. The authors recognized that the hump could be either the signature of a second smaller gap or a consequence of errors in $C_{lat}$.

### B. Ba$_{0.6}$K$_{0.4}$Fe$_2$As$_2$: Ch. Kant et al.[3]

The specific heat of a sample with $T_c = 37.3$ K was measured from 2 to 300 K. Both $C_{lat}$ and $\gamma_n$ were obtained by fitting the normal-state data above $T_c$ as $C = \gamma_n T + C_D(\theta_D, T) + 2C_E(\theta_{E1}, T) + 2C_E(\theta_{E2}, T)$ where $C_D$ and $C_E$ are Debye and Einstein functions with characteristic temperatures $\theta_D$, $\theta_{E1}$, and $\theta_{E2}$. These are the correct numbers of Debye and Einstein terms for this material but to limit the number of adjustable parameters only two of the Einstein temperatures were independently adjusted in the fit. The fit gave $\gamma_n = 49$ mJ K$^{-2}$ mol$^{-1}$ and the effective value of B$_3$ = 0.651 mJ K$^{-4}$ mol$^{-1}$, the coefficient of the $T^3$ term in the Debye function, as given by Eq. (4). The low-temperature data were not analyzed to obtain a value of B$_3$. The derived $C_{es}$ gave an entropy at $T_c$ that satisfied the entropy-conservation condition to within 1.6%. It was fitted with a single-gap $\alpha$-model expression, giving $\alpha = 2.07$. A two-gap fit with $\alpha_1$ and $\alpha_2$ fixed at 3.7 and 1.9 (from ARPES data[21]) gave $\gamma_{n1} = 9.2$ mJ K$^{-2}$ mol$^{-1}$ and $\gamma_{n2} = 39.8$ mJ K$^{-2}$ mol$^{-1}$, but it was concluded that because of the extra parameters the two-gap fit was "not superior" to the single-gap fit.

### C. Ba$_{0.68}$K$_{0.32}$Fe$_2$As$_2$: P. Popovich et al.[39]

The specific heat of a sample with $T_c \sim 38.3$ K was measured from 2 to 200 K. Both $C_{lat}$ and $\gamma_n$ were obtained by fitting the normal-state data between 40 and 150 K as $C = \gamma_n T + C_{lat}$, with $C_{lat}$ taken to be $C_{lat}$ for a Mn-doped sample scaled by two adjustable parameters. The fit, made "under the constraint of entropy conservation", gave $\gamma_n = 50$ mJ K$^{-2}$ mol$^{-1}$ and the scaling parameters in $C_{lat}$. A low-temperature fit gave $\gamma_r = 1.2$ mJ K$^{-2}$ mol$^{-1}$. The derived $C_{es}/T$ shows a conspicuous "knee" at $\sim 15$ K, which is well represented by a two-gap $\alpha$−model fit with $\gamma_{n1} \sim \gamma_{n2} \sim 25$ mJ K$^{-2}$ mol$^{-1}$, $\alpha_1 = 3.3$, and $\alpha_2 = 1.1$. However, the effective value of B$_3$ obtained in the fit and used in calculating $C_{es}$, 0.462 mJ K$^{-4}$ mol$^{-1}$, differs significantly from the correct value, 0.496 mJ K$^{-4}$ mol$^{-1}$, which was obtained in a low-$T$ fit to data for the K doped sample. This discrepancy itself would lead to an overestimate of 7.7 mJ K$^{-2}$ mol$^{-1}$ in $C_{es}/T$ at 15 K, which is comparable to the magnitude of the knee.

### D. Ba$_{0.65}$Na$_{0.35}$Fe$_2$As$_2$: Pramanik et al.[40]

Specific-heat data for a sample with $T_c = 29.4$ K are shown for $1.8 − 35$ K. The "estimated" value of $\gamma_n$, 57.5 mJ K$^{-2}$ mol$^{-1}$, includes a residual $\gamma_r = 3.3$ mJ K$^{-2}$ mol$^{-1}$, but its origin is not specified. $C_{lat}$ was taken to be $C_{lat}$ for the undoped orthorhombic BaFe$_2$As$_2$ scaled by the factor 0.95, which was chosen to give entropy conservation for the derived $C_{es}$ with the estimated $\gamma_n$. The value of B$_3$ in the derived $C_{lat}$, i.e., the value reported for BaFe$_2$As$_2$ scaled by the factor 0.95, is 0.35 mJ K$^{-2}$ mol$^{-1}$, but the correct value, as estimated from the inset in Fig. 2, is $\sim 0.72$ mJ K$^{-4}$ mol$^{-1}$. The difference between these numbers would produce an overestimate of $C_{es}/T$ of $\sim 0.37T^2$ mJ K$^{-2}$ mol$^{-1}$ at the lowest temperatures, and below $\sim 7$ K $C_{es}/T$ has approximately this form. The derived $C_{es}$ shows conspicuous deviations from BCS behavior, positive for t ≤ 0.6 and negative for t ≥ 0.6. A two-gap $\alpha$-model fit gave $\gamma_{n1}$ = 29.9 mJ K$^{-2}$ mol$^{-1}$, $\gamma_{n2}$ = 27.6 mJ K$^{-2}$ mol$^{-1}$, $\alpha_1$ = 2.08, and $\alpha_2$ = 1.06.

### E. Ba$_{0.55}$K$_{0.45}$Fe$_{1.95}$ Co$_{0.05}$As$_2$: Gofryk et al.[30]

The specific heat of a sample with $T_c$ = 32.5 K was measured from 1.8 to 300 K. Both $C_{lat}$ and $\gamma_n$ were obtained by fitting the normal-state data as $C = \gamma_n T + (1-k)C_D(\theta_D, T) + kC_E(\theta_E, T)$. $C_D$ and $C_E$ were said to be Debye and Einstein functions, but this would not account for the correct number of phonon modes, leaving some ambiguity about the nature of the fitting expression. The fit gave $\gamma_n$ = 40.5 mJ K$^{-2}$ mol$^{-1}$. With allowance for a residual $\gamma_r$ = 2.24 mJ K$^{-2}$ mol$^{-1}$, the fit satisfied the entropy-conservation requirement. A single-gap $\alpha$-model fit gave $\alpha$ = 2.57. A two-gap fit, which was a better representation of $C_{es}$, gave $\gamma_{n1}$ = 34.8 mJ K$^{-2}$ mol$^{-1}$, $\gamma_{n2}$ = 5.7 mJ K$^{-2}$ mol$^{-1}$, $\alpha_1$ = 3.9, and $\alpha_2$ = 0.86.

The variety of methods used to obtain $C_{lat}$ in just these five examples suggests a general awareness of the problems in separating the lattice contribution, even though they are not mentioned specifically. For four of the five the specific-heat results were analyzed in two successive, but interdependent, fits. In the initial fit, normal-state data above $T_c$ were fitted as the sum of $\gamma_n T$ and an expression for $C_{lat}$. In A $\gamma_n$ had been determined independently and the fit gave $C_{lat}$; in B, C, and E, the fit was used to obtain $\gamma_n$, as well as $C_{lat}$. The $C_{lat}$ derived in the initial fit was then extrapolated to below $T_c$ to obtain $C_{es}$, which was compared with $\alpha$-model expressions in the second fit to obtain the parameters characteristic of the individual bands. Since the $\alpha$ model relates all thermodynamic properties with thermodynamic consistency, the second fit gives plausible results only if $C_{es}$ and $\gamma_n$ are at least approximately consistent with entropy conservation. In each case entropy conservation was satisfied, but this result was a consequence of the initial fit, which had given the $C_{lat}$ that led to a $C_{es}$ consistent with $\gamma_n$ and entropy conservation. Uncertainties in the reported values of $\gamma_n$ are suggested by the method of their determination in A and D, by the question of the validity of the high-$T$ fits in B, C, and E, and by the effect of satisfaction of the entropy-conservation requirement (see below) in all five. Uncertainties in the low-$T$ $C_{lat}$, and in the resultant low-$T$ $C_{es}$, are suggested by the extrapolations from higher temperatures in A, B, C, and E, by its derivation in D, and by their dependence on the values of $\gamma_n$ in all five.

To investigate the nature of the errors that can result from the high-$T$ fits that are made to determine $C_{lat}$ we fit our normal-state data as $C = \gamma_n T + C_D(\theta_D, T) + 2C_E(\theta_{E1}, T) + 2C_E(\theta_{E2}, T)$, the same fitting expression used in B. Good fits to the data, with rms deviations ~ 0.5%, were obtained for four different fitting intervals ─ 40 to 100, 40 to 150, 40 to 250, and 40 to 300 K ─ *but they gave four different values of $\gamma_n$*, ranging from 73 to 92 mJ K$^{-2}$ mol$^{-1}$. Although this fitting expression provides an adequate representation of the relatively weak $T$ dependence of $C/T$, the parameters are not uniquely determined by the fit. Depending on the $T$ interval of the fit, the derived parameters change as necessary to compensate for inadequacies in the $T$ dependences of the various terms. The four fits gave small differences in the values of $\theta_{E1}$ and $\theta_{E2}$, which are more important at higher temperatures, but more significant differences in the values of $\theta_D$, which is critically important at the lowest temperatures where it determines the effective value of $B_3$, as given by Eq. (4). Extrapolations of these results for $C_{lat}$ to low temperatures gave impossible results for $C_{es}$: For the fits that gave $\gamma_n = 73$ and 92 mJ K$^{-2}$ mol$^{-1}$, the values of $B_3$ were 1.34 and 1.06 mJ K$^{-4}$, whereas the correct value is 0.602 mJ K$^{-2}$ mol$^{-4}$. As a consequence, the derived values of $C_{es}$ were *negative* below ~ 21 K, and the entropies at $T_c$ differed from $\gamma_n T_c$ by factors of ~ 3. The $\gamma_n$ and $C_{lat}$ derived in these fits are clearly not correct. It is instructive to compare the 40 to 300 K fit to our data, a free fit that did not give entropy conservation, with the 40 to 300 K fit to data for a similar sample in B that gave entropy-conservation to ~ 1.6%: Our fit gave $\gamma_n = 77.5$ mJ K$^{-2}$ mol$^{-1}$ and a negative $C_{es}$; the fit in B gave $\gamma_n = 49$ mJ K$^{-2}$ mol$^{-1}$ and a $C_{es}$ that was still negative in the vicinity of 8 K, but more than an order of magnitude smaller than that obtained from our fit. Evidently the latitude in these high-$T$ fits allows substantial effects of the entropy-conservation constraint that may take the form of reducing gross errors in the derived $\gamma_n$ and $C_{lat}$. Overall, these results emphasize the possibility of substantial errors in the values of $\gamma_n$ obtained in the high-$T$ fits. Imposition of the entropy-conservation requirement may reduce the error in $\gamma_n$, without ensuring high accuracy, and it does not preclude significant errors in the resulting $C_{es}$ (see Sec. II

Four of the other five values of $\gamma_n$ are substantially higher than ours and only one, in E, is reasonably close (see Table III). It is not possible to estimate the uncertainties in the values of $\gamma_n$ related to the details of the high temperature fits, but there are some points that merit comment. In A, the very high value of $\gamma_n$ was determined by extrapolating $\gamma_v(H)$ to a high value of $H_{c2}$ that does not seem to have been borne out by subsequent measurements. In D, the origin of the intermediate value of $\gamma_n$ is not specified. These values are substantially higher than ours on average, but, given their wide variation and the differences in their derivation, they do not constitute evidence that ours is in error.

The parameters characteristic of the individual bands obtained in the $\alpha$-model fits to $C_{es}$ are sensitive to the details of the $T$ dependence of $C_{es}$. Approximate satisfaction of the entropy-conservation requirement may eliminate gross errors in $C_{es}$, but it does not preclude small $T$-dependent errors that can affect the values of the parameters obtained in the fits. Parameters characteristic of the large-gap band are sensitive to details of $C_{es}$ at higher temperatures and the errors in $C_{lat}$ and $\gamma_n$ discussed above. Parameters characteristic of the small-gap band are more sensitive to the details of the $T$ dependence of $C_{es}$ at lower temperatures and particularly to errors in the effective value of $B_3$ in the $C_{lat}$ that was used to derive $C_{es}$. In A, the "hump" in $C_{es}$ near 20 K seems to be a consequence of the high value of $\gamma_n$, which ensures high *average* values of $C_{es}$, while use of the correct value of $B_3$ in $C_{lat}$ shifts the high values to higher temperatures. The hump is probably an indication of an error in $C_{lat}$ rather than evidence of a second band with a small gap, and the authors recognized that possibility. In B, the two-gap fit was judged to be not superior to the single-gap fit but the data do suggest the presence of a small-gap band. In C, the difference between the effective value of $B_3$ in $C_{lat}$ and the correct value would produce a contribution to $C_{es}$ comparable to that attributed to a small-gap band, which suggests some uncertainty in the magnitude of the reported contribution of a small-gap band. In D, the contribution to $C_{es}$ attributed to a small-gap band seems to be at least partly a $T^2$ contribution introduced by the difference between the effective value of $B_3$ in $C_{lat}$ and the correct value. In E, the authors note that there is some discrepancy between the experimental data and the two gap fit that gives the values of the parameters characteristic of the two bands, but the fit does represent the data reasonably well. The values of the parameters characteristic of the individual bands reported in E could be said to be in reasonable agreement with ours but with that exception the other values are all substantially different from ours (see Table III). However, given the wide ranges of the values of the other parameters, the differences with our values do not constitute evidence of errors in ours.

## VII. SUMMARY

The specific heat of a high-quality single crystal of $Ba_{0.59}K_{0.41}Fe_2As_2$, a near-optimally hole-doped superconductor in the 122 series of Fe pnictides, was measured from 2 to 300 K, and below 50 K in fields to $\mu_0 H = 14$ T.

A novel method of analysis of the data, based on *direct* comparisons of $\alpha$-model expressions for the electron contribution with the *total* measured specific heat, was used to obtain the parameters characteristic of two electron bands. The parameters characteristic of a small-gap band were obtained in an analysis of the specific-heat data below 12 K, where the contribution of the large-gap band is negligible. The parameters characteristic of a large-gap band were obtained from the discontinuities in $C$ and $dC/dT$ at $T_c$ after correcting for the contributions of the small-gap band. The total DOS, as measured by the value of $\gamma_n$, 40.1 mJ K$^{-2}$ mol$^{-1}$, is the sum of two contributions, $\gamma_{n1} = 31.0$ mJ K$^{-2}$ mol$^{-1}$ and $\gamma_{n2} = 9.1$ mJ K$^{-2}$ mol$^{-1}$, from bands with superconducting-state energy gaps that are, respectively, larger and smaller than the weak-coupling BCS value. In terms of the parameter $\alpha$, which is $\alpha_{BCS} = 1.764$ in the weak-coupling limit of the BCS theory, $\alpha_1 = 3.30$ and $\alpha_2 = 0.86$. The energy gaps derived from the specific-heat data are within the ranges of values obtained in ARPES measurements, but there are some significant differences. The $H$ dependence of the $T$-proportional term in the vortex-state specific heat suggests a nodeless order parameter and is consistent with extended s-wave pairing. The relations between the DOS and energy gaps for the two bands are not consistent with theoretical predictions[37] for a model in which superconductivity is produced by interband interactions alone. Comparison of the total DOS, as deduced from the value of $\gamma_n$, with band-structure calculations shows a strong effective mass renormalization that is without precedent in similar materials and is not theoretically explained.

The analysis bypasses the independent determination of the lattice contribution, an essential step in the conventional analyses in which the lattice contribution is subtracted from the total to obtain the electron contribution, which is then compared with the $\alpha$-model expressions. It eliminates the substantial uncertainties in the electron contribution associated with the approximations inherent in the determination of the lattice contribution. The parameters characteristic of the electron contribution are significantly different from those obtained by conventional analyses for five other near-optimally hole-doped $BaFe_2As_2$ superconductors. The parameters obtained in the conventional analyses differ significantly among themselves, which could be a consequence of the different approximations used for the lattice contribution.

## ACKNOWLEDGEMENTS

This work was supported by the Director, Office of Science, Office of Basic Energy Sciences, U.S. Department of Energy, under Contract No. DE-AC02-05CH11231 and Office of Basic Energy Sciences U.S. DOE under Grant No. DE-AC03-76SF008. We are grateful to J. E. Gordon for help with the $\alpha$-model calculations and helpful discussions about the method of analyzing the data.

[*] Corresponding author: rotundu@stabford.edu

[a] CRR and TRF contributed equally to this work.

[b] Current address: Stanford Institute for Materials and Energy Sciences, SLAC National Accelerator Laboratory, 2575 Sand Hill Road, Menlo Park, California 94025, USA

[c] Current address: European Synchrotron Radiation Facility, BP 220, F-38043 Grenoble Cedex, France

**TABLES**

TABLE I. Characteristic parameters of the two electron bands.

| Electron band | $\alpha$ | $\Delta(0)$ (meV) | $\gamma$ (mJ K$^{-2}$ mol$^{-1}$) |
|---|---|---|---|
| 1 | 3.30 | 10.49 | 9.1 |
| 2 | 0.86 | 2.73 | 31.0 |

TABLE II. The $H$-dependent parameters derived in a "global" fit with Eq. (12) to the data for $2 \leq T \leq 12$ K in 10 fields, $0 \leq \mu_0 H \leq 14$ T. The fit gave $\gamma_v(H)$, $a(H)$, and $b(H)$ directly, and it also gave the parameters $\theta_{Sch}(0)$ and $\beta$ that determine $\theta_{Sch}(H)$.

| $\mu_0 H$ (T) | $\gamma_v$ (mJ K$^{-2}$ mol$^{-1}$) | $\theta_{Sch}$ (K) | $a$ (mJ K$^{-1}$ mol$^{-1}$) | $b$ |
|---|---|---|---|---|
| 0 | $0.00 \pm 0.025$ | 7.32 | $337. \pm 17.$ | $1.00 \pm 0.04$ |
| 0.5 | $0.48 \pm 0.085$ | 7.35 | $281. \pm 19.$ | $0.90 \pm 0.04$ |
| 1 | $1.07 \pm 0.086$ | 7.44 | $253. \pm 20.$ | $0.87 \pm 0.04$ |
| 2 | $2.02 \pm 0.088$ | 7.79 | $220. \pm 20.$ | $0.82 \pm 0.03$ |
| 4 | $3.54 \pm 0.119$ | 9.06 | $172. \pm 20.$ | $0.74 \pm 0.03$ |
| 6 | $5.03 \pm 0.182$ | 10.84 | $143. \pm 21.$ | $0.67 \pm 0.03$ |
| 8 | $6.57 \pm 0.202$ | 12.94 | $114. \pm 21.$ | $0.60 \pm 0.02$ |
| 10 | $7.89 \pm 0.169$ | 15.21 | $95. \pm 20.$ | $0.53 \pm 0.03$ |
| 12 | $8.93 \pm 0.196$ | 17.60 | $79. \pm 16.$ | $0.43 \pm 0.05$ |
| 14 | $9.75 \pm 0.278$ | 20.05 | $84. \pm 14.$ | $0.39 \pm 0.04$ |

TABLE III. Characteristic parameters of the electron bands, as derived from six different measurements. The values in the top row, this work, were derived by comparing $\alpha$-model expressions for the electron contribution directly with the total measured specific heat. The values in rows A to E were derived in conventional analyses in which the $\alpha$-model expressions were compared with a superconducting-state electron specific heat that had been obtained by subtracting an independently determined approximation for the lattice contribution from the total specific heat. The values of $\gamma_n$ are the totals for two bands, however they were derived; the values of $\alpha$ are the results of single-band fits, if they were made; the values in the 4$^{th}$ – 7$^{th}$ columns are the results of two-band fits. For the two-band fit in B $\alpha_1$ and $\alpha_2$ were fixed at values obtained from ARPES measurements. The units of $\gamma_n$, $\gamma_{n1}$, and $\gamma_{n2}$ are mJ K$^{-2}$ mol$^{-1}$.

| Reference | $\gamma_n$ | $\alpha$ | $\gamma_{n1}$ | $\gamma_{n2}$ | $\alpha_1$ | $\alpha_2$ |
|---|---|---|---|---|---|---|
| This work | 40.1 | | 31.0 | 9.1 | 3.30 | 0.86 |
| A | 71. | 1.9 | | | | |
| B | 49.0 | 2.07 | 9.2 | 39.8 | 3.7 | 1.9 |
| C | 50. | | 25. | 25. | 3.3 | 1.1 |
| D | 57.5 | | 29.9 | 27.6 | 2.08 | 1.06 |
| E | 40.5 | 2.57 | 34.8 | 5.7 | 3.9 | 0.86 |

**FIGURES**

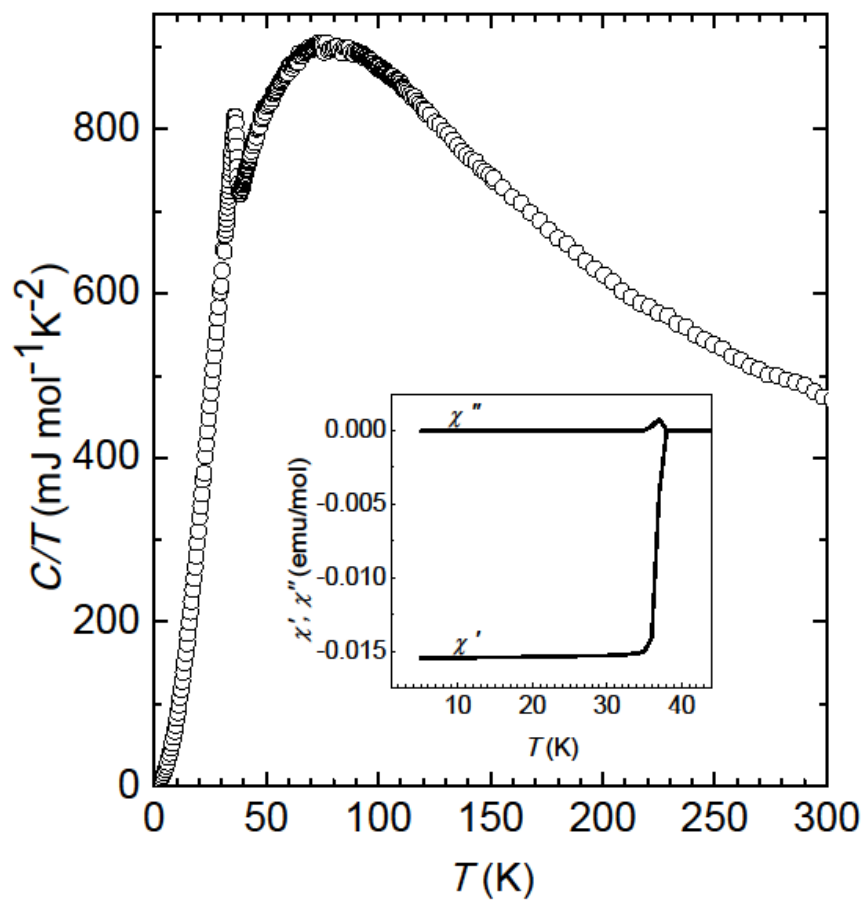

FIG. 1. The specific heat in zero field, as *C/T* *vs* *T*, for 2 - 300 K. The superconducting transition is marked by the sharp peak in *C/T* near $T_c$ = 36.9 K. At that temperature, as shown in the inset, the ac susceptibility shows a sharp and complete transition to the superconducting state.

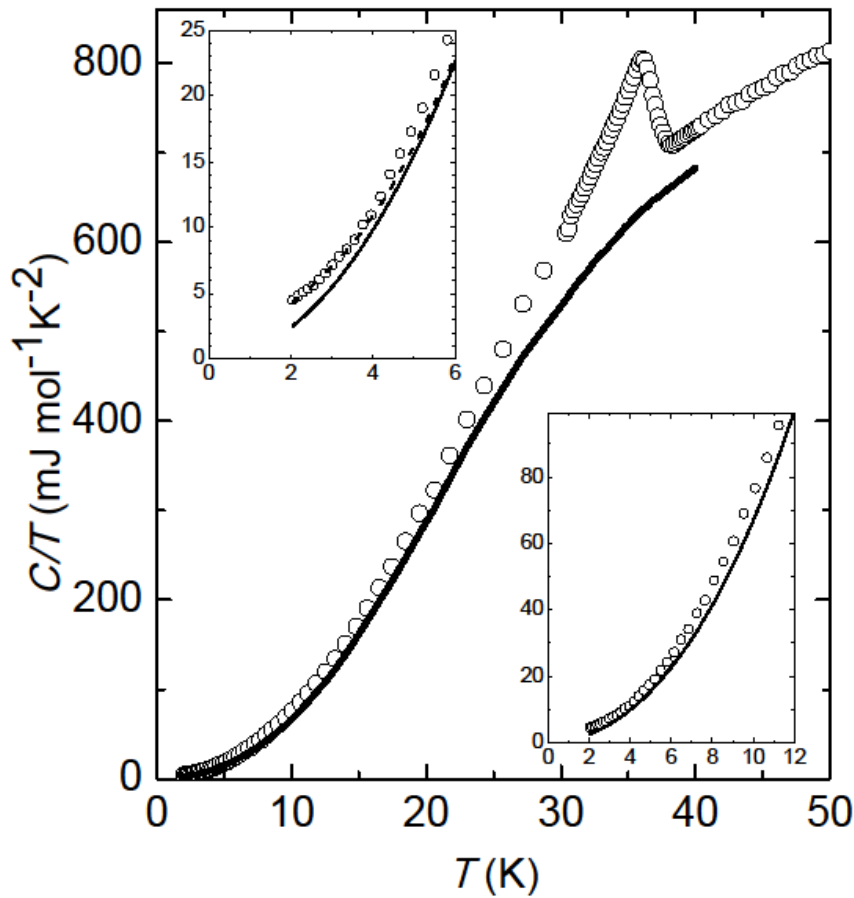

FIG. 2. The specific heat in zero field, as $C/T$ *vs* $T$, for 2 - 50 K in the main panel, and for intervals at lower temperatures in the insets. The superconducting transition is marked by the discontinuity in $C$ near $T_c = 36.9$ K. The solid curves represent the apparent $C_{lat}$, obtained by different methods above and below 12 K as described in the text. The dashed curve in the upper inset represents $C_{lat} + C_{Sch}$ in zero field, as determined in a "global" fit to the data for $2 \leq T \leq 12$ K in 10 fields, $0 \leq \mu_0 H \leq 14$ T.

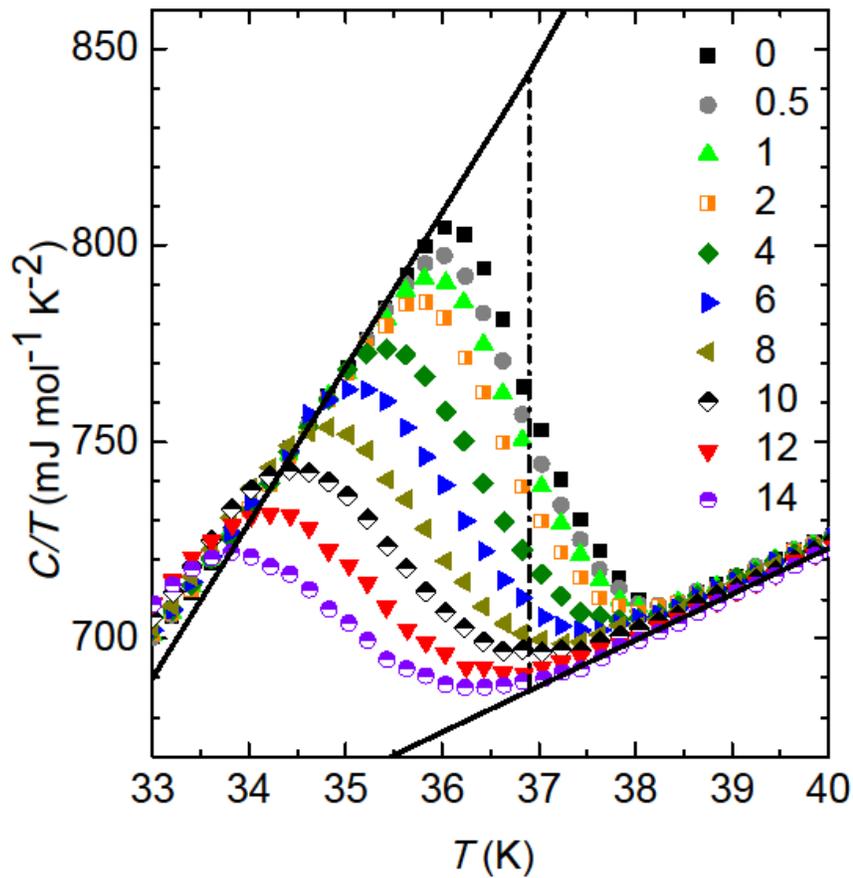

FIG. 3. (Color on line) The specific heat in the vicinity of $T_c$, as $C/T$ *vs* $T$. The solid sloping lines are the results of fits to the data just outside the transition region (see text). The dashed, vertical line is an entropy-conserving construction that determines $T_c$ as 36.9 K. The extrapolations of the solid lines to $T_c$ represent the mean-field transition in zero field.

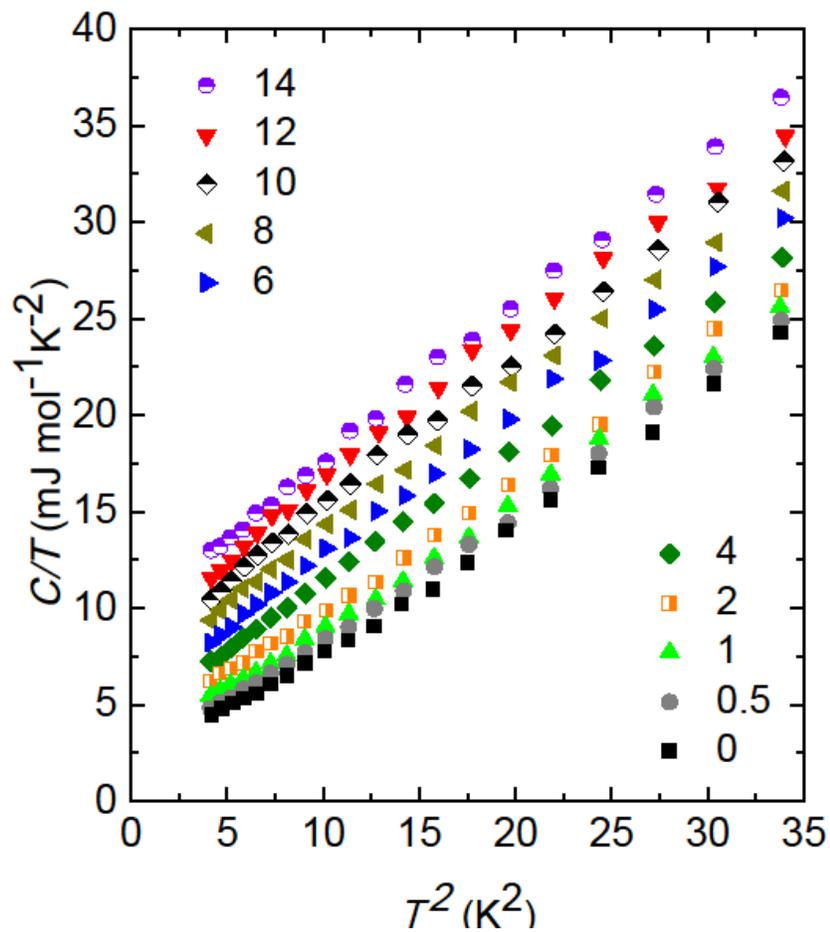

FIG. 4. (Color on line) The specific heat as $C/T$ vs $T^2$ to 6 K in 10 fields, $0 \leq \mu_0 H \leq 14$ T. The deviations from linearity suggest the presence of a low concentration of paramagnetic centers.

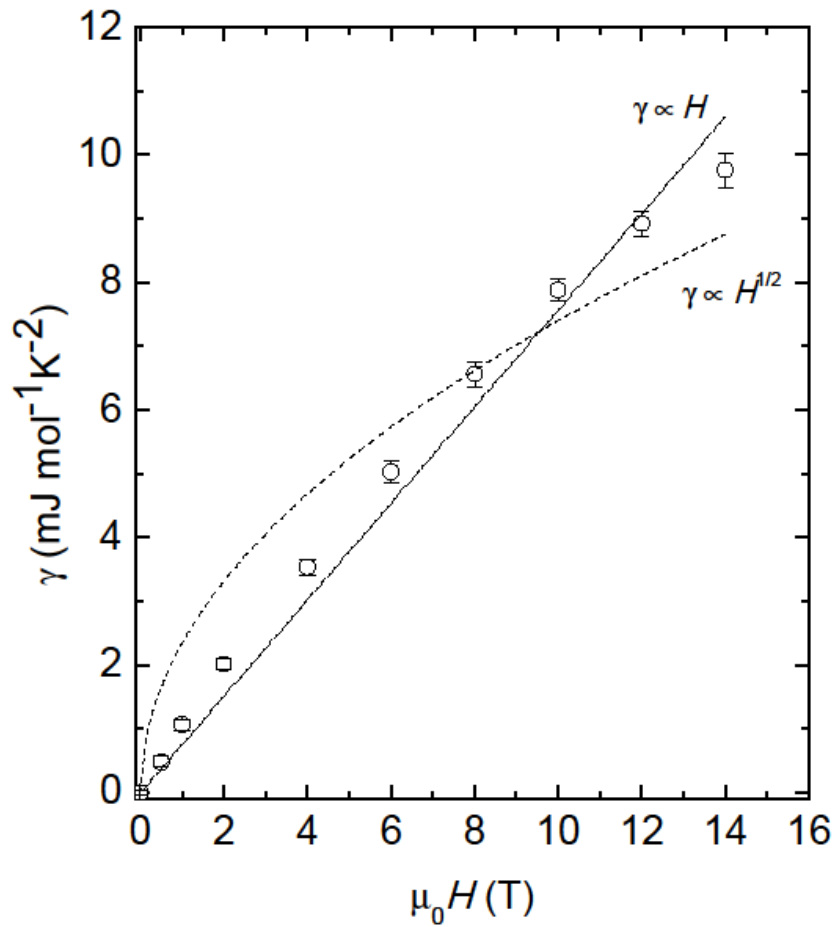

FIG. 5. The $H$ dependence of $\gamma_V(H)$ as obtained in a "global" fit to the data for $2 \leq T \leq 12$ K in 10 fields, $0 \leq \mu_0 H \leq 14$ T. The solid and dashed lines represent least-squares fits to $H$ and $H^{1/2}$ dependences (see text). The error bars correspond to the uncertainties determined in the fit.

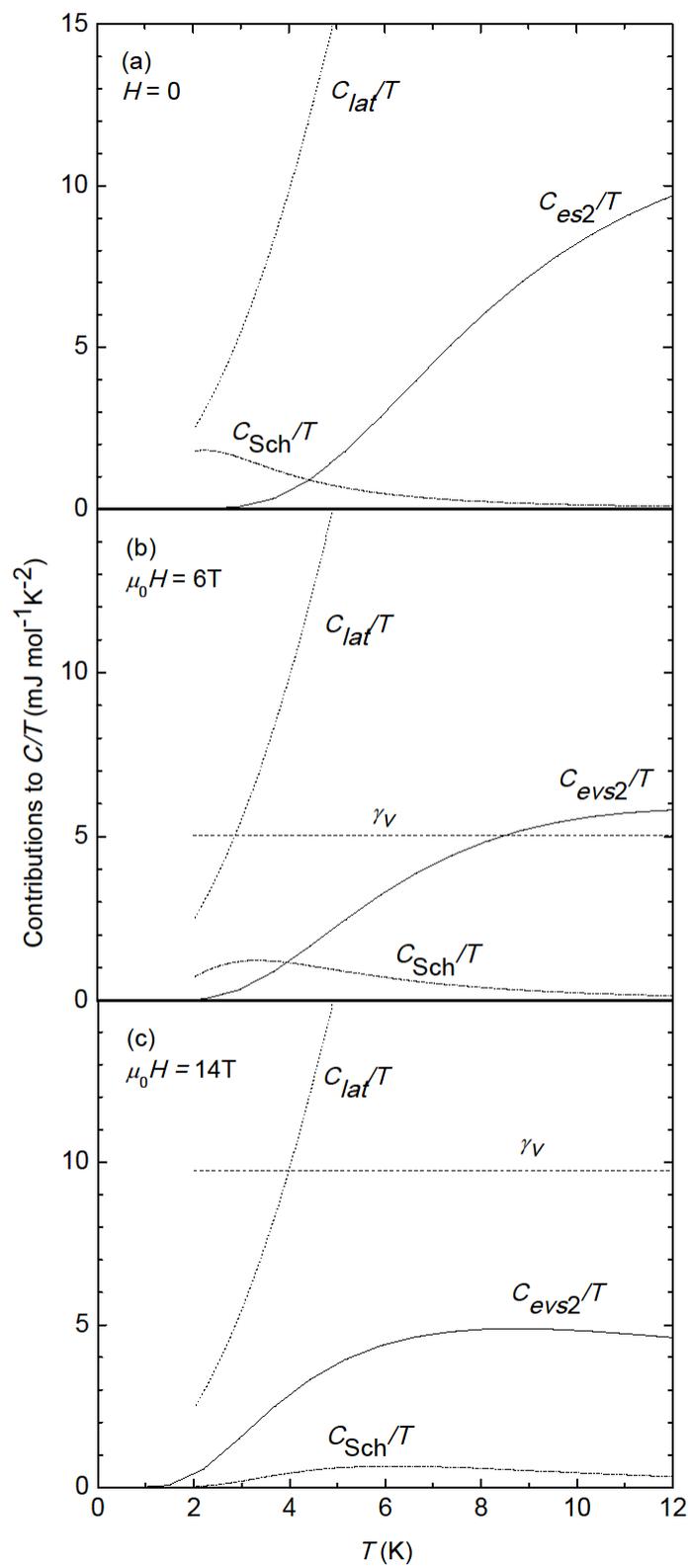

FIG. 6. Lattice, paramagnetic-center, and electron contributions to C/T, for $\mu_0 H = 0$, 6, and 14 T in (a), (b), and (c), as obtained in a "global" fit to the data for $2 \leq T \leq 12$ K in 10 fields, $0 \leq \mu_0 H \leq 14$ T. In (a) $C_{es2}/T$ is the contribution of the small-gap band to $C_{es}/T$, i.e., in the superconducting state. In (b) and (c) $C_{evs2}/T$ is the corresponding contribution of the small-gap band to $C_{ev}/T$, i.e., in the vortex state. In this temperature interval and on this scale the analogous contributions of the large-gap band are negligible. In (b) and (c) $\gamma_v$ is the total contribution of the vortex cores.

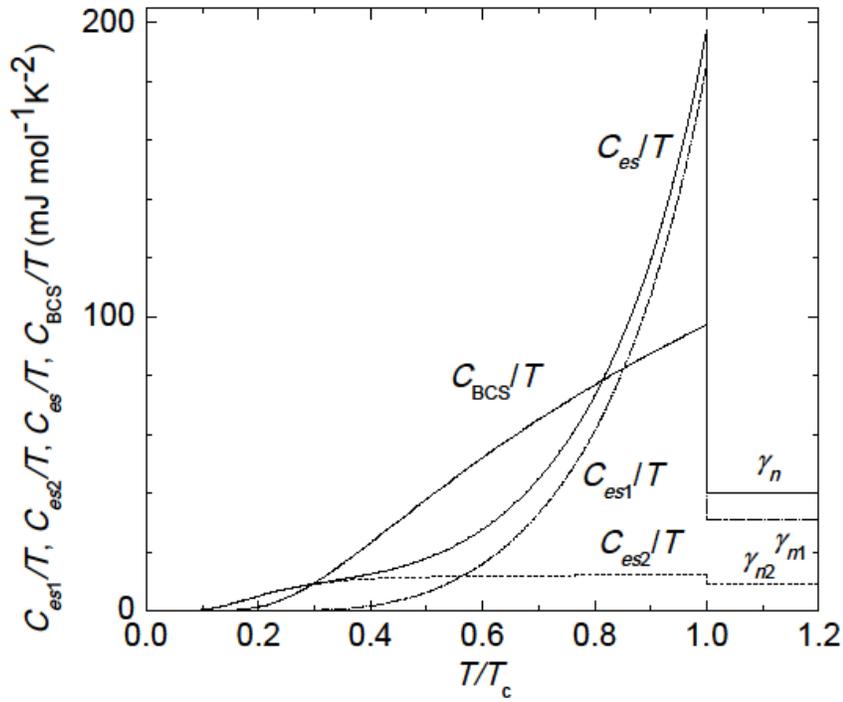

FIG. 7. The electron contribution to C/T and its two components as functions of $T/T_c$, in the superconducting state, for $T/T_c \leq 1$, and in the normal state, for $T/T_c \geq 1$. The dash/dot lines labeled $C_{es1}/T$ and $\gamma_{n1}$ are the large-gap band component; the dashed lines labeled $C_{es2}$ and $\gamma_{n2}$ are the small-gap band component; the solid lines labeled $C_{es}/T$ and $\gamma_n$ are their sum. The dotted line labeled $C_{BCS}/T$ is the result of the BCS theory in the weak-coupling limit for the same $\gamma_n$.